\documentclass[twocolumn,showpacs,preprintnumbers,amsmath,amssymb]{revtex4}
\usepackage{graphicx}
\usepackage{bm}
\def\PTP{Prog. Theor. Phys.(Kyoto)}
\def\NPA{{Nucl. Phys.} {\bf A}}
\def\NPB{{Nucl. Phys.} {\bf B}}
\def\PLB{{Phys. Lett.} B}

\def\PRD{{Phys. Rev.} D}

\begin{document}
\newcommand{\ttbs}{\char'134}


\title{Infrared Features of the  Landau Gauge QCD }

\author{Sadataka Furui}
\email{furui@umb.teikyo-u.ac.jp}
\homepage{http://albert.umb.teikyo-u.ac.jp/furui_lab/furuipbs.htm}
\affiliation{%
School of Science and Engineering, Teikyo University, 320-8551 Japan.
}%

\author{Hideo Nakajima}
\email{nakajima@is.utsunomiya-u.ac.jp}
 
\affiliation{
Department of Information Science, Utsunomiya University, 321-8585 Japan. 
}%

\date{29 October 2003}

\begin{abstract}
The infrared features of Landau gauge QCD are studied by  lattice
simulation of $\beta=6.0,\ 16^4,\ 24^4,\ 32^4$ and $\beta=6.4,\ 32^4,\ 48^4$.
We adopt two definitions of the gauge field; (1) $U$ linear and (2) $\log U$, and
measured the gluon propagator and ghost propagator. The infrared singularity of the gluon propagator is less than that of the tree level result, but the gluon propagator at 0 momentum remains finite. The infrared singularity of the ghost propagator is stronger than the tree level.
The QCD running coupling measured by using the gluon propagator and the 
ghost propagator  has a maximum  $\alpha_s(p)\simeq 1$ at around $p=0.5$ GeV and decreases as $p$ approaches 0. The data are analyzed with use of the formula of the principle of minimal sensitivity and the effective charge method, and by the contour-improved perturbation method, which suggest the necessity of the resummation
of the perturbation series in the infrared region together with the existence of 
the infrared fixed point. 
The Kugo-Ojima parameter is about -0.8 in contrast with the theoretically
expected value of -1. 
The color off-diagonal  part of the ghost propagator in the Landau gauge is  consistent with zero, and its fluctuation can be parametrized as a constant$/(qa)^4$.

\end{abstract}

\maketitle

\section{\label{sec:level1}Introduction}
Two decades ago, Kugo and Ojima proposed a criterion for the absence
of colored massless asymptoptic states in Landau gauge QCD using the Becchi-Rouet-Stora-Tyutin(BRST)
symmetry \cite{KO}. They suggested to measure the two point function of the
covariant derivative of the ghost and the commutator of the antighost and gauge field
\begin{eqnarray}
&&(\delta_{\mu\nu}-{q_\mu q_\nu\over q^2})u^{ab}(q^2)\nonumber\\
&&={1\over V}
\sum_{x,y} e^{-ip(x-y)}\langle  {\rm tr}\left({\lambda^a}^{\dag}
D_\mu \displaystyle{1\over -\partial D}[A_\nu,\lambda^b] \right)_{xy}\rangle.
\end{eqnarray}

They showed that $u^{ab}(0)=\delta^{ab}u(0)$ satisfies
\begin{equation}
1+u(0)=\frac{Z_1}{Z_3}=\frac{1}{\tilde Z_3},
\end{equation}
where $Z_3$ is the gluon wave function renormalization factor, $Z_1$ is the gluon vertex renormalization factor, and $\tilde Z_3$ is the ghost wave function renormalization factor, respectively.  Kugo claimed that confinement is realized either by
(1) $Z_1=0$ and $Z_3=$finite or (2) $Z_1/Z_3=0$ but(and) $Z_3=0$.
The divergence of $\tilde Z_3$ implies $u(0)=-1$ and the presence of a long-range
correlation between colored sources.

As shown by Gribov \cite{Gr}, the Landau gauge is not unique and the uniqueness of the gauge field can be achieved via restriction to the fundamental modular region (FMR), i.e. the region where the norm is the absolute minimum \cite{Zw,Zw1}.  We adopted the smearing gauge fixing \cite{HdF} to make the configuration close to the FMR. We observed  proximity of the gauge configurations with and without  smearing gauge fixing, but the overlap of the Gribov region and FMR remains to be investigated.

The confinement scenario was recently reviewed in the framework of the renormalization 
group equation and dispersion relation \cite{Kond}. It was shown that the gluon dressing function $Z_A(q^2)$ defined from the gluon propagator of $SU(n)$ 
\begin{eqnarray}
D_{\mu\nu}(q)&=&{1\over n^2-1}\sum_{x={\bf x},t}e^{-ikx}Tr\langle A_\mu(x)^\dagger A_\nu(0)\rangle \nonumber\\
&=&(\delta_{\mu\nu}-{q_\mu q_\nu\over q^2})D_A(q^2),
\end{eqnarray}
as $Z_A(q^2)=q^2 D_A(q^2)$,
satisfies the superconvergence relation, and the gluon dressing function at zero momentum does not necessarily vanish as Gribov and Zwanziger conjectured, but it could be finite. A systematic study of lattice data indeed establishes the infrared
finiteness of the gluon propagator \cite{Bon}.

The ghost propagator is defined as the Fourier transform of an expectation value of the inverse Faddeev-Popov(FP) operator $\cal  M$
\begin{equation}
D_G^{ab}(x,y)=\langle {\rm tr} \langle \lambda^a x|({\cal  M}[U])^{-1}|
\lambda^b y\rangle \rangle,
\end{equation}
where the outmost $\langle\cdots \rangle$ denotes average over samples $U$. 
The infrared behavior of the ghost propagator 
in the renormalization group approach depends on the gauge, and whether it satisfies the superconvergence relation is not clear. 
In maximal abelian gauge, it is conjectured that the off-diagonal gluon and off-diagonal ghost satisfy the superconvergence relation, but the diagonal ghost does not, and it is the source of the long-range correlation.  We remark that Nishijima proposed a sufficient condition for the confinement as $Z_3=0$, based on the convergence of the spectral function \cite{Ni}. 
 
The non-perturbative color confinement mechanism was studied with the Dyson-Schwinger approach \cite{SHA} and lattice simulations \cite{Cu, adelade, orsay1, NF,FN,scgt,lat03}. We produced $SU(3)$ gauge configurations by using the heat-bath method \cite{CaMa,KePe} and performed gauge fixing \cite{NF}.
We analyzed lattice Landau gauge configurations of $\beta=6.0$, $16^4,24^4,32^4$and $\beta=6.4$, $32^4,48^4$ produced at KEK. Progress reports are presented
in \cite{NF} and an extensive report will be published elsewhere.
The gauge field is defined from the link variables as
\begin{itemize}
\item $\log U$ type:
$U_{x,\mu}=e^{A_{x,\mu}},\ A_{x,\mu}^{\dag}=-A_{x,\mu},$
\item $U$ linear type:
$A_{x,\mu}=\displaystyle{1\over 2}(U_{x,\mu}-U_{x,\mu}^{\dag})|_{trless\ p.}$
\end{itemize}
The  fundamental modular region of lattice size $L$ is specified by the
 global minimum along the gauge orbits, i.e., 

$\Lambda_L=\{U|F_{U}(1)={\rm Min}_gF_{U}(g)\}$, 
$\Lambda_L\subset \Omega_L$, 
where $\Omega_L$ is called the  Gribov region (local minima) and \\
$\Omega_L=\{U|-\partial { D(U)}\ge 0\ ,\ \partial A(U)=0\}.$

Here $F_U(g)$ for the two options are \cite{MN,SF}
\begin{itemize}
\item 
$F_U(g)=||A^g||^2=\sum_{x,\mu}{\rm tr}
 \left({{A^g}_{x,\mu}}^{\dag}A^g_{x,\mu}\right)$,
\item
$F_U(g)=\sum_{x,\mu}\left (1- {1\over 3}{\rm Re}\ {\rm tr}U^g_{x,\mu}\right),$
\end{itemize}
respectively.

The Landau gauge fixing in the $\log U$ type is performed by Newton's method where the linear equation is solved up to third order of the gauge field, and then the Poisson equation
is solved by the Green's function method for $16^4, 24^4$ lattices and 
by the multigrid method for $32^4, 48^4$ lattices \cite{NF}.
 The gauge fixing in the $U$ linear
type is performed by the standard overrelaxation method. The  accuracy of $\partial A(U)=0$ is
$10^{-4}$ in the maximum norm.

In the calculation of the ghost propagator, i.e. inverse FP operator, we adopt 
the perturbative method with use of the multigrid Poisson solver \cite{Hack}, 
whose accuracy was kept within $10^{-5}$,
and we set 1\% as an ending condition in the method \cite{NF}. But later we
also introduce the straightforward and preconditioned conjugate gradient(CG) methods \cite{Ort} for cross-checking of the calculation. In the preconditioned CG method, we take for the preconditioning operation the same truncated perturbation series of inversion as that of the perturbative method. 
 In the CG method, the accuracy of the
convergence of the series is set to be less than 5\% in the $L_2$ norm.

We analyze these data using a method inspired by the principle of minimal sensitivity (PMS) and/or the effective charge method \cite{PMS,Gru} and the contour-improved perturbation method \cite{HoMa}.

In sec. II we explain the method of analysis, and in sec. III the
lattice data are presented and compared with results of the theoretical analysis. We performed a cross-check of our program of SU(3) by performing the SU(2)
lattice simulation. In sec. IV we show our data and compare with those results of other groups. Our conclusion and outlook are presented in sec. V.  Some details of our method of calculating the FP inverse operator are given in the appendix.

\section{Method of analysis}

In the infrared region, the QCD perturbation series does not converge and 
truncation of the renormalization group equation and resummation of the
series to evaluate the renormalon effect was proposed \cite{tHo}. On the
other hand, the possibility of the presence of an infrared fixed point was
discussed and methods to bridge infrared and ultraviolet regions
via the renormalization group equation were proposed \cite{PMS,Gru}. The method was
recently applied to an analysis of lattice data \cite{vAc} and succeeded in explaining qualitatively the data. We briefly review the method.
  
\subsection{PMS and the effective charge method}

In the PMS method, the $n$th-order approximation to the physical quantity ${\cal R}$ is expressed by the corresponding series of coupling constant $h^{(n)}$
which is  defined as a solution of
\begin{eqnarray}\label{heq}
\beta_0\log\frac{\mu^2}{\Lambda^2}=\frac{1}{h}+\frac{\beta_1}{\beta_0} \log(\beta_0 h)\qquad\qquad\qquad\qquad\qquad\nonumber\\
+\int_0^h dx (\frac{1}{x^2}-\frac{\beta_1}{\beta_0 x}-\frac{\beta_0}{\beta_0 x^2+\beta_1 x^3+\cdots+\beta_n x^{n+2}}),
\end{eqnarray}
where the scheme-independent constant and logarithmic term are separated.

 When ${\cal R}$ is the QCD running coupling from the triple gluon vertex from up to three-loop diagrams in the modified minimal subtraction ($\overline{\rm MS}$) scheme, one sets the scale $\mu^2$ equal to the external scale $q^2$ and expresses
\begin{equation}
{\cal R}^n=h^{(n)}(1+A_1h^{(n)}+A_2h^{(n)2}+\cdots+A_nh^{(n)n})\label{req}
\end{equation}
where in the case of $n=3$, $A_1=70/3$, $\displaystyle A_2=\frac{516217}{576}-\zeta_3\frac{153}{4}$,
$\displaystyle A_3=\frac{304676635}{6912}-\zeta_3\frac{299961}{64}-\zeta_5\frac{81825}{64}$ \cite{ChRe}.

When one defines  $y_{\overline{\rm MS}}(q)$ as a solution of
\begin{equation}
1/y_{\overline{\rm MS}}(q)={\beta_0 \log(q/\Lambda_{\overline{\rm MS}})^2-\frac{\beta_1}{\beta_0}\log(\beta_0 y_{\overline{\rm MS}}(q))}
\end{equation}
and expresses the solution of Eq. (\ref{heq})
\begin{eqnarray}
h(q)&=&y_{\overline{\rm MS}}(q)\{ 1+y_{\overline{\rm MS}}(q)^2(\bar\beta_2/\beta_0-(\beta_1/\beta_0)^2)\nonumber\\
&+&y_{\overline{\rm MS}}(q)^3\frac{1}{2}(\bar \beta_3/\beta_0-(\beta_1/\beta_0)^3)+\cdots\},
\end{eqnarray}
where $\beta_0=11, \beta_1=102$ ,
$\displaystyle\bar\beta_2=\frac{2857}{2}$, $\displaystyle\bar\beta_3=\frac{149753}{6}+3564\zeta_3$, we can calculate ${\cal R}$ via eq.(\ref{req}).

The parameter $y_{\overline{\rm MS}}(q)$ can be expressed as $y$ defined as a solution of 
\begin{equation}
 \beta_0\log\frac{\mu^2}{\Lambda^2}=\frac{1}{y}+\frac{\beta_1}{\beta_0}\log(\beta_0 y)
\end{equation}
and the function
\begin{equation}
 k(q^2,y)=\frac{1}{y}+\frac{\beta_1}{\beta_0}\log(\beta_0 y)-\beta_0\log(q^2/\Lambda_{\overline{\rm MS}}^2).
\end{equation}

In \cite{vAc} the parameter $y$ is fixed via
minimization of $|({\cal R}^n(y)-{\cal R}^1(y))/{\cal R}^1(y)|$ for each $q^2$. There are subtle problems in fixing $y$ of PMS in the low-energy region \cite{BroLu},We leave the fitting of the low energy region for the future and we fix $y$ at  $\mu=1.97$ GeV by solving
\begin{equation}
1/y={\beta_0 \log(\mu/\Lambda)^2-\frac{\beta_1}{\beta_0}\log(\beta_0 y)}.
\end{equation}
The choice of $\mu=1.97$ GeV corresponds to the inverse lattice unit $1/a$ of
$\beta=6.0$ and chosen by \cite{orsay1} as the factorization scale of the effective charge method.
 When $\Lambda=\Lambda_{\overline{\rm MS}}=0.237$ GeV we find the solution $y=0.01594$, and we call this method of choosing $y$ at a specific $\mu$ and define $\alpha_s$ from ghost-ghost-gluon coupling, the $\widetilde{{\rm MOM}}$ scheme. 

\subsection{Contour-improved perturbation series}

Exact solution of the two-loop renormalization group equation for $x$ with variable $\displaystyle t=\log \frac{q^2}{\Lambda^2}$ 
\begin{equation}
 \beta(x)=\frac{dx}{dt}=-\frac{b}{2} x^2(1+cx), 
\end{equation}
is 
\begin{equation}
 \frac{b}{2} \log(q^2/\Lambda^2)=\frac{1}{x}-c\log(\frac{1}{x}+c).
\end{equation}
The solution $x$ can be expressed as
$\displaystyle x(q^2)=-\frac{1}{c}\frac{1}{1+W(z)}$, where $W(z)$ is the Lambert W function which satisfies $W(z)e^{W(z)}=z$. We apply the dispersion relation and consider contributions on a cut of negative real axis in the space of $q^2$, i.e. take $q$ pure imaginary. In order to be consistent with the $\overline{\rm MS}$ scheme, the variable $z$ is defined as 
\begin{eqnarray}
z&=&-e^{(-1-b t/2c)}=-\frac{1}{e}(\frac{q}{\tilde\Lambda_{\overline{\rm MS}}})^{-b/c}e^{iK\pi}\nonumber\\
&=&-Z(q^2)e^{iK\pi},
\end{eqnarray}
where $t=\log (q^2/\tilde\Lambda_{\overline{\rm MS}}^2)$, $\tilde\Lambda_{\overline{\rm MS}}=(2c/b)^{-c/b}\Lambda_{\overline{\rm MS}}$, $K=-b/2c$ \cite{PMS,HoMa}. The physical quantities ${\cal R}$ are expressed in a series
\begin{equation}
{\cal R}(q^2)={\cal B}_1(q^2)+\sum_{n=1}^\infty A_n{\cal B}_{n+1}(q^2)
\end{equation}
\begin{equation}
{\cal B}_n=\frac{1}{2\pi}\int_{-\pi}^\pi (\frac{-1}{c(1+W(Z(q^2)e^{iK\theta})})^n d\theta .
\end{equation}

\section{Lattice data}
\subsection{Gluon propagator}

\begin{figure}
\begin{center}
\includegraphics{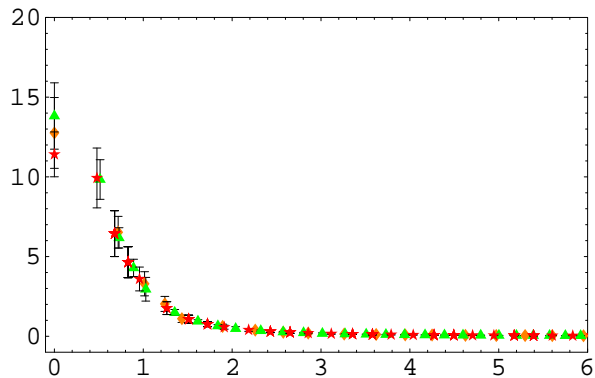}
\end{center}
\caption{The gluon propagator as the function of the momentum $q$(GeV). $\beta=6.0$, $24^4$(triangles), $32^4$(diamonds) and $\beta=6.4$, $48^4$(stars) in $\log U$ definition. }\label{gl243248}
\end{figure}

The gluon propagator on the lattice was measured by using cylindrical cut method \cite{adelade}, i.e., choosing momenta close to the diagonal direction. When the difference of their lattice constant $a^{-1}=1.885$ GeV in $\beta=6.0, 32^3\times 64$ and our $a^{-1}=1.97$ GeV, $32^4$ is taken into account, the data are consistent with \cite{adelade}(see Fig. \ref{gl243248}).

The effective coupling $y$ of the $\widetilde{{\rm MOM}}$ scheme is 
calculated from 
\begin{equation}
1/y={\beta_0 \log(\mu/\Lambda_z)^2-\frac{\beta_1}{\beta_0}\log(\beta_0 y)}\label{ygluon}\end{equation}
for $\mu=1.97$ GeV and  $\Lambda_z=\Lambda_{\overline{\rm MS}}e^{25085/37752}$ \cite{ChRe} obtained from the three-gluon vertex in Landau gauge perturbation theory.   The relevant solution of eq.(\ref{ygluon}) is  $y=0.02227$. 

\begin{figure}
\begin{center}
\includegraphics{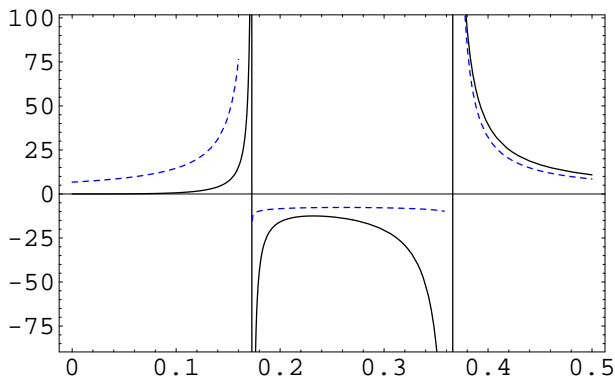}
\end{center}
\caption{ The dressing function $\lambda_z Z(q^2,y)$ as a function of the variable $z$ for a fixed $y=0.02227$. The branches of the Lambert $W$ funtion $1/W_0(0.17-z)$, Re$[W_{-1}(0.17-z)+W_{-1}(z-e^{-1})]$, and $1/W_0(z-e^{-1})$ are shown as dotted lines.}\label{zinv}
\end{figure}

The gluon dressing function is defined as $Z_A(q^2)=q^2 D_A(q^2)$. Its inverse
$Z^{-1}$ is expressed in the two-loop perturbation series as \cite{ChRe}
\begin{eqnarray}
 Z^{-1}(q^2,y)&=&\lambda_z^{-1} h^{(2)-13/22}(1-\frac{25085}{2904}h^{(2)}\nonumber\\
&-&(\frac{41245993}{1874048}-\frac{9747}{352}\zeta_3)h^{(2)2}),
\end{eqnarray}
where $\lambda_z$ is a fitting parameter(see Fig. \ref{zinv}). 

\begin{figure}
\begin{center}
\includegraphics{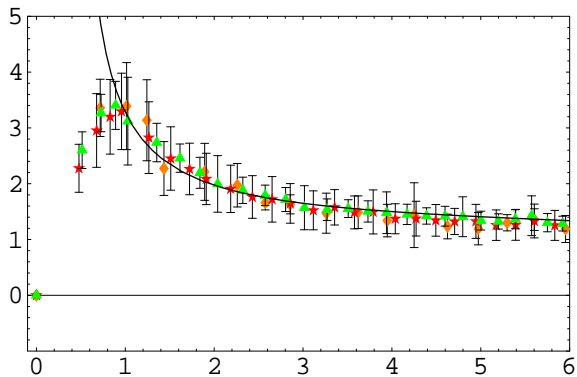}
\end{center}
\caption{The gluon dressing function as the function of the momentum $q$(GeV). $\beta=6.0$, $24^4$(triangles), $32^4$(diamonds) and $\beta=6.4$, $48^4$(stars) in the $\log U$ version. The fitted line is that of the $\widetilde{\rm MOM}$ scheme.}\label{dgl243248}
\end{figure}

As shown in Fig. \ref{gl243248} and Fig. \ref{dgl243248}, the gluon propagators of $24^4, 32^4$ and $48^4$ as a function of the physical momentum agree quite well with one another and they can be fitted by
\begin{equation}
D_A(q^2)=\frac{Z(q^2,y)|_{y=0.02227}}{q^2}=\frac{Z_A(q^2)}{q^2}
\end{equation}
in the $q>0.8$ GeV region.
At zero momentum,  $D_A(0)$ decreases as the lattice size becomes larger.

The gluon dressing function in the $\widetilde{\rm MOM}$ scheme with $y=0.02227$ fits the lattice data for $q>0.8$ GeV, but there appears a discontinuity at $z\simeq 0.174$ and $z=1/e$.@We note that the $q$ dependence of $Z(q^2,y)$ in $z<0.17$ is similar to ${1}/{W_0(0.17-z)}$, that in $0.17< z <1/e $ is similar to Re$[W_{-1}(0.17-z)+W_{-1}(z-e^{-1})]$, and that in $z>1/e$ is similar to ${1}/{W_0(z-1/e)}$. 

\subsection{Ghost propagator}

The ghost dressing function is defined by the ghost propagator as $G^{ab}(q^2)=q^2 {D_G}^{ab}(q^2)$. 
In the $\widetilde{\rm MOM}$ scheme, we fix the scale by choosing  $y$ as a solution of
\begin{equation}
1/y={\beta_0 \log(\mu/\Lambda_{gh})^2-\frac{\beta_1}{\beta_0}\log(\beta_0 y)}.
\end{equation}
For  $\mu=1.97$ GeV and $\Lambda_{gh}=\Lambda_{\overline{\rm MS}}e^{1757/2904}$ \cite{ChRe} obtained from two-loop Landau gauge perturbation theory, we find as a relevant solution  $y=0.02142$.

The ghost dressing function is
\begin{eqnarray}
&&{Z_g}^{-1}(q^2,y)={\lambda_g^{-1}}h^{(2) -9/44}[1+h^{(2)}(-\frac{5271}{1936})\nonumber\\
&+&h^{(2) 2}(-\frac{615512003}{7496192}+\frac{5697}{704}\zeta_3)+\cdots],
\end{eqnarray}
where $\lambda_g$ is a fitting parameter.

\begin{figure}[htb]
\begin{center}
\includegraphics{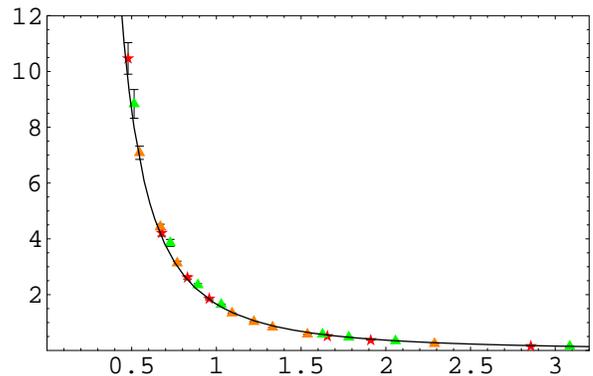}
\end{center}
\caption{The ghost propagator as the function of the momentum $q$(GeV). $\beta=6.0$, $24^4, 32^4$ and $\beta=6.4$, $48^4$ in the $\log U$ version. The fitted line is that of the $\widetilde{\rm MOM}$ scheme. $Z_g(q^2,y)$ is singular at $\tilde\Lambda_{\overline{\rm MS}}\simeq 0.25$ GeV which should be washed away by the nonperturbative effects. }\label{gh243248}
\end{figure}

In Fig. \ref{gh243248}, 
the $24^4, 32^4$, and $48^4$ lattice data are compared with 
\begin{equation}\label{dg}
D_G(q^2)=-\frac{Z_g(q^2,y)|_{y=0.02142}}{q^2}=\frac{G(q^2)}{q^2}.
\end{equation}

We observe that the agreement is good for $q>0.5$ GeV and better than the result of the PMS method of \cite{vAc}. 
The ghost propagators were calculated by the perturbative method and the straightforward and preconditioned CG methods.
We found that the two CG methods are consistent and give better accuracy than the perturbative method in SU(2),  and give correct result in the lowest momentum point of SU(3) $48^4$ lattice. With the lowest momentum point of the $48^4$ lattice calculated with the CG method,  the whole data can be fitted by Eq.(\ref{dg}).

\subsection{QCD running coupling}

We measured the running coupling from the product of the gluon dressing function and the ghost dressing function squared \cite{SHA}.
\begin{equation}
\alpha_s(q^2)=\frac{g_0^2}{4\pi}Z_A(q^2){ G(q^2)}^2\simeq (qa)^{-2(\alpha_D+2\alpha_G)}.
\end{equation}

The lattice size dependences of the exponents $\alpha_D$ and $\alpha_G$ are summarized in Table \ref{exp}.
\begin{table}[htb]
\caption{ The exponent of gluon dressing function near zero momentum  $\alpha_D$, near $qa=1$  $\alpha_D'$, and the exponent of the ghost dressing function near zero momentum $\alpha_G$.  $\log U$ type.}\label{exp}
\begin{tabular}{c|c|ccc|c}
 $\beta$ &$L$ &{$\alpha_D$ } & {$\alpha_D'$ } & {$\alpha_G$} &{$\alpha_D+2\alpha_G$}\\
\hline
6.0 &{32}& {-0.375}  & {0.302} & {0.174}&{-0.03(10)} \\
6.4 &{48}& {-0.273}  & {0.288} & {0.193} &{0.11(10)}\\
\hline
\end{tabular}
\end{table}

\begin{figure}[htb]
\begin{center}
\includegraphics{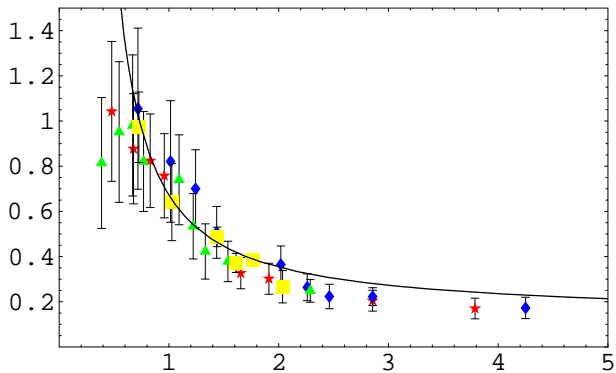}
\end{center}
\caption{The running coupling $\alpha_s(q)$ of $\beta=6.0$, $24^4$(squares), 
$32^4$(triangles), $\beta=6.4$, $32^4$(diamonds) and $48^4$(stars) 
as a function of momentum $q$(GeV) and the result of the 
PMS method in the $\widetilde{\rm MOM}$ scheme.} \label{alp3248}
\end{figure}

The effective running coupling in the $\overline{\rm MS}$ scheme is expressed by
the series of coupling constants $h^{(n)}$ as eq.(\ref{req}) \cite{vAc,ChRe}.
The result of the $\widetilde{{\rm MOM}}$ scheme using $y=0.01594$ is shown by the solid line in Fig. \ref{alp3248}. The lattice data of $24^4, 32^4$ and $48^4$ and the $\widetilde{\rm MOM}$ scheme agree in 0.5GeV$<q<$2GeV, but the fit is slightly overestimated at $q>2$ GeV.

\begin{figure}[htb]
\begin{center}
\includegraphics{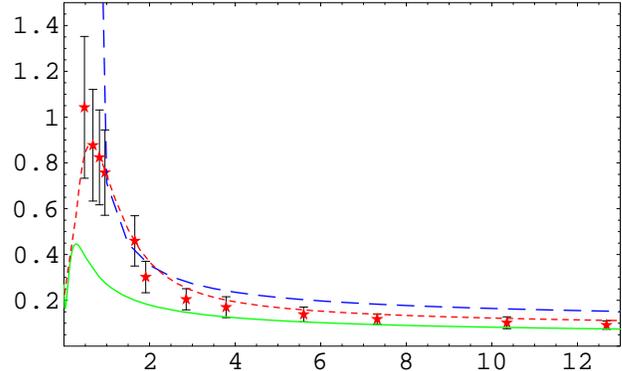}
\end{center}
\caption{The running coupling $\alpha_s(q)$ as a function of momentum $q$(GeV) of the $\beta=6.4, 48^4$ lattice. The solid line is the result of ${\cal R}^2$ using $\Lambda_{\overline{\rm MS}}=237$ MeV. Dotted line is 
the result of $e^{70/6\beta_0}\tilde\Lambda_{\overline{\rm MS}}$ and including half of $A_2$.
The dashed line is the result of the $\widetilde{\rm MOM}$ scheme. 
}\label{alppert1}
\end{figure}

In the contour-improved perturbation series, the running coupling in two loops is expressed as
\begin{equation}
{\cal R}^2(q^2)={\cal B}_1(q^2)+A_1{\cal B}_2(q^2)+A_2{\cal B}_3(q^2)+\cdots.
\end{equation}
The series truncated at order $A_1$ 
is plotted in Fig. \ref{alppert1} together with the $48^4$ lattice data
measured by using the $\log U$ definition. 
We observed that the fit is good for $q>2 $GeV, but the non-perturbative effect is underestimated. 

Since in the perturbative calculation of the Landau gauge gluon vertex in the $\overline{\rm MS}$ scheme, the $\Lambda_{\overline{\rm MS}}$ is modified to $e^{70/6\beta_0}\Lambda_{\overline{\rm MS}}$ \cite{CeGo}, we performed the same replacement. The result overestimates in  $q>1$ GeV region. Since the $A_3$ term is not known and there are cancellations between sucessive terms, we fit the data by inclusion of half of the $A_2$ term. The result is shown by the dotted line in Fig. \ref{alppert1}.

The similar nonperturbative effect was attributed to the gluon condensates in \cite{orsay1,orsay2}. The lattice data are qualitatively the same as the
results of hypothetical $\tau$ lepton decay \cite{Bro}, and the Dyson-Schwinger approach  \cite{FAR}.

The lowest-momentum point of $\alpha_s$ of $\beta=6.4, 48^4$ becomes consistent with 
results of other lattice sizes when it is calculated with the CG method. 

\subsection{Kugo-Ojima parameter}

Our lattice data of (1) the Kugo-Ojima parameter $c=-u(0)$, (2)the trace of the gauge field divided by the dimension $e/d$, and (3) the deviation parameter $h$ from the horizon condition  \cite{NF} are summarized in Table \ref{kugotab}.  
\begin{figure}[htb]
\begin{center}
\includegraphics{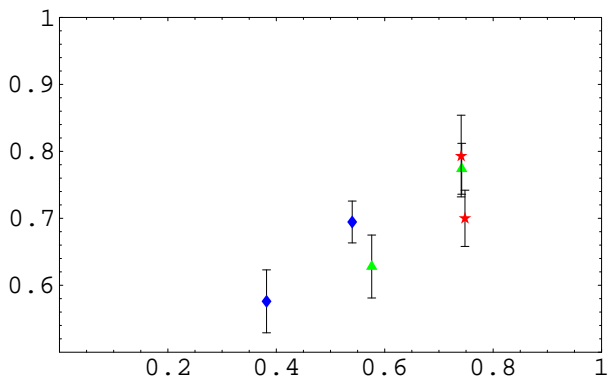}
\caption{The Kugo-Ojima parameter $c$ as a function of $\log Z$(1.97GeV).  $\beta=6.4$,  $32^4,\ 48^4$ in $\log U$(stars),  $\beta=6.0$, $16^4$ and $24^4$ in $\log U$(triangles) and $U$ linear (diamonds) versions.}
\label{kg32}
\end{center}
\end{figure}

We observe that the Kugo-Ojima parameter of the $U$ linear definition remains smaller than that of $\log U$. The similar difference exists in the ghost propagator in the infrared region.

\begin{table}[htb]
\caption{The Kugo-Ojima parameter $c$ in  the $U$ linear and $\log U$  versions. $\beta=6.0$ and $6.4$.}\label{kugotab}
\begin{center}
\begin{tabular}{c|c|ccc|ccc}
 $\beta$&$L$ &$c_1$ & $e_1/d$ & $h_1$ & $c_2$ & $e_2/d$ & $h_2$ \\
\hline
6.0 &16&  0.576(79) &   0.860(1) & -0.28 & 0.628(94)& 0.943(1) & -0.32\\
6.0 &24&  0.695(63)  &  0.861(1) & -0.17 & 0.774(76)& 0.944(1) & -0.17\\
6.0 &32&  0.706(39)  &  0.862(1) & -0.15 & 0.777(46)& 0.944(1) & -0.16\\
\hline
6.4 &32& 0.650(39) & 0.883(1) & -0.23 & 0.700(42)& 0.953(1) & -0.25\\
6.4 &48&           &          &       & 0.793(61)& 0.954(1) & -0.16\\
\end{tabular}
\end{center}
\end{table}

We plot in Fig. \ref{kg32} the value $c$  as a function of $\log Z(\mu^2)$ of $\beta=6.0$, $16^4$, and $24^4$ in $\log U$ and $U$ linear definitions and $\beta=6.4$, $32^4$ and $48^4$ in the $\log U$ definition. The value $c$ of $\beta=6.0$, $32^4$ is almost the same as $24^4$. 
 The value increases as the lattice size increases from $16^4$ to $24^4$ and the extrapolation of the two definitions to those of a large lattice where $c$ in $\log U$ and $U$ linear seem to cross at $c\sim 1$. The linear extrapolation as the function of $\log Z(\mu^2)$ is based on the factorizablity
\begin{equation}
Z_3(\mu^2,\Lambda_{\widetilde{\rm MOM}})=Z_R(\mu^2)/Z_b(\Lambda_{\widetilde{\rm MOM}})
\end{equation}
when $\mu\sim 1.97$ GeV, which allows us to express
\begin{eqnarray}
Z_3(\mu^2,\Lambda_{\widetilde{\rm MOM}})&=&Z_3(\mu^2,\Lambda_{\overline {\rm MS}})\nonumber\\
&\times&({Z_b}^{-1}(\Lambda_{\widetilde{\rm MOM}})/{Z_b}^{-1}(\Lambda_{\overline {\rm MS}})).
\end{eqnarray}
 The difference of the speed of $Z_b(\Lambda_{\widetilde{\rm MOM}})$ to its continuum limit in the $U$ linear and $\log U$ definitions will appear as a difference of the slope. However, the increase of $c$ from $24^4$ to $32^4$ is small. The Kugo-Ojima parameter $c$ of $\beta=6.4, 48^4$ lattice  calculated  in the CG method is $0.793(61)$, which is consistent to the
result of the $\log U$ definition of $\beta=6.0,\ 24^4$, and $32^4$ lattice data.

\section{SU(2) lattice data}

In the numerical simulation of the SU(2) Yang-Mills field, we took the $U$ linear type
gauge field and simulated  $\beta=2.2$ and $16^4$ lattices.  We took 67 samples taken after 18 000 thermalization sweeps and up to 84 000 sweeps with intervals of 1000 sweeps \cite{lat03}. To each sample, we performed parallel tempering gauge fixing (PT) and direct gauge fixing by the overrelaxation method (first copy). 
We define the scale by the relation $1/a=0.938$ GeV and compare our data with those of  \cite{BCLM,AFF} and  \cite{cuc}.

\subsection{Gluon propagator}
\begin{figure}[htb]
\begin{center}
\includegraphics{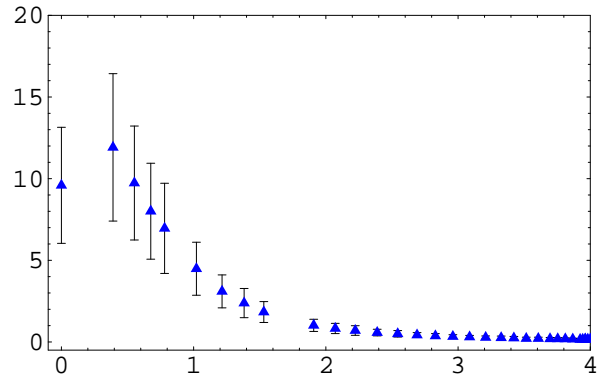}
\end{center}
\caption{The gluon propagator $D_A(q)$  as a function of the  momentum $q$(GeV) of PT samples.  }\label{glsu2}
\end{figure}

The gluon propagator is shown in Fig. \ref{glsu2}.  We observe that above 1 GeV our data agree with  \cite{BCLM}, but in the infrared region our data have an enhancement. Suppression at 0 momentum is 
consistent with the data of  \cite{AFF}.

\subsection{Ghost propagator}
\begin{figure}[htb]
\begin{center}
\includegraphics{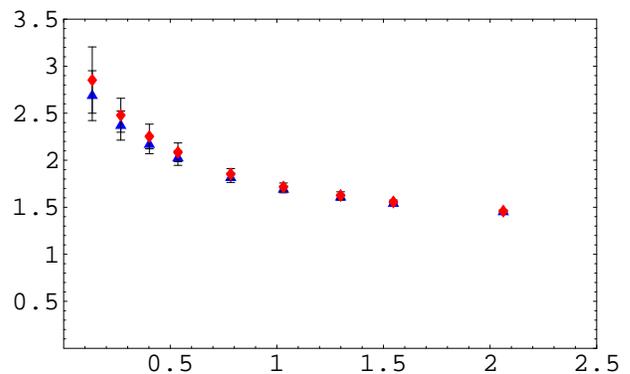}
\end{center}
\caption{The color diagonal ghost dressing function $D_G(q^2)\times q^2$  as a function of the momentum squared $q^2$(GeV$^2)$.  The first copies (diamonds) are more singular than PT (triangles).  }\label{ghsu2}
\end{figure}

The color diagonal component of the ghost propagator calculated in PT
is about $6\%$ less singular than that of first copy (Fig. \ref{ghsu2}).
We performed the calculation of the FP inverse operator by using the CG method, since the matrix is symmetric positive definite. Our data in the infrared are
 less singular than that of  \cite{BCLM}. Although
there are difference in the gauge fixing method (PT versus simulated annealing), we do not understand the origin of the difference.

In the maximal abelian (MA) gauge, color symmetry is spontaneously broken by the 
ghost condensation  \cite{KoShi, Sha}. In the Landau gauge, there is no background field as in the MA gauge, but the structure of the color off-diagonal ghost propagator has not been known. In order to investigate this problem, we measured color off-diagonal symmetric and antisymmetric ($\epsilon_{abc}D_G^{ab}(q,q)$) matrix elements, where $D_G^{ab}(q,q)$ is the ghost propagator with color indices $a$ and $b$.  We observed that the color
off-diagonal antisymmetric part is consistent with zero pointwise as is
expected from the  theoretical observation and that the color
off-diagonal symmetric part multiplied 
by $q^4$ is consistent with zero over the ensemble average, but its
standard 
deviation is almost constant in the whole momentum region.
 The fluctuation can be parametrized as $\sigma/q^4$ with $\sigma=0.0176(28)$ GeV$^2$, in the normalization tr $\tau^2=1$. We observed the same qualitative features in the SU(3) $48^4$ lattice, but $\sigma$ is about 1/9 of the SU(2) $16^4$ lattice, i.e., the fluctuation is statistical.

\subsection{QCD running coupling}
\begin{figure}[htb]
\begin{center}
\includegraphics{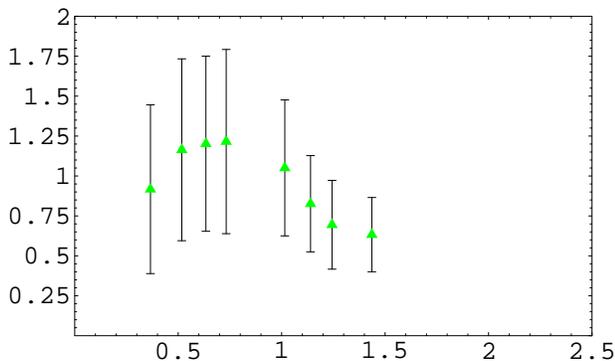}
\end{center}
\caption{The running coupling $\alpha_s(q)$  as a function of the  momentum $q$(GeV) of PT samples.  }\label{alpsu2}
\end{figure}

The result of the running is shown in Fig.\ref{alpsu2}. As a result of the difference in the ghost propagator, the running coupling is about 1/3 of  \cite{BCLM}. We observe suppression near 0 momentum.

\subsection{Kugo-Ojima parameter}
The Kugo-Ojima parameter $c$ of the PT samples was 0.690(52) and that of the first copy was 0.722(68). This difference is qualitatively the same as that of the ghost dressing function at 0 momentum.

\section{Conclusion and outlook}
There are two aspects of color confinement: i.e., (1) the presence of long-range correlation between colored sources and (2) the absence of massless gluon poles.
The Kugo-Ojima criterion is a sufficient condition for the two aspects, but the
lattice data do not verify that these criteria are  satisfied. 

A new method of FMR gauge fixing in SU(2) is reported in \cite{lat03}.
We observe that the gluon propagator is suppressed at zero momentum in SU(2),
i.e., the exponent $\alpha_D <-0.5$, in contrast to the SU(3) case where
$\alpha_D\geq -0.5$. In the simulation of SU(2), we observed differences in the Kugo-Ojima parameter of the configuration in the FMR and of copies randomly produced in the Gribov region.
The Gribov copy affects the Kugo-Ojima parameter, and in the ghost propagator in
the infrared region, the difference is about 4\%. 
 Color SU(3) contains I, U and V SU(2) spin components, and we expect that the Gribov ambiguity is the same order.

In the lattice data, the singularity of the ghost propagator is stronger than the tree level and that of the gluon propagator is weaker than the tree level. The dependence on the $U$ linear or $\log U$ definition of the gauge field is small in the gluon propagator consistent with  \cite{BM}, but not negligible in the FP inverse operator. 

We aimed at detecting in the lattice dynamics a signal of the Kugo-Ojima confinement
 criterion derived in the continuum theory, formulated with use of the FP Lagrangian and BRST symmetry. We also noted that Zwanziger's horizon condition, based on the lattice formulation, coincides  with the Kugo-Ojima criterion  \cite{Zw,NF}. However, our present data  are not satisfactory to prove or disprove the confinement criterion. 
The color off-diagonal antisymmetric part of the ghost propagator \cite{dudal,KoShi} vanishes in the Landau gauge,  but the off-diagonal symmetric part has fluctuations proportional to $(qa)^{-4}$.

Although there are problems in fixing $y$ of the PMS in the low-energy region, an extension of the effective charge method is
a possible solution. In an extension of the solution of the two-loop renormalization 
group equation expressed by the Lambert-$W$ function, a solution of Pad\'e approximant of the three-loop renormalization group equation was shown \cite{GGK} and numerical
calculation was done for $N_f\geq 3$ \cite{Mag}. In the analytical perturbation theory
approach in one loop, one predicts  \cite{Shir} a non-perturbative infrared fixed point
of $\displaystyle \frac{\alpha_s(0)}{4\pi}=\frac{1}{\beta_0}=\frac{1}{11-2/3 N_f}$. Extension to the two-loops is discussed in  \cite{GGK}.
For $N_f=0$, one needs  continuation.  There is a conjecture, in
combination with the conformal relation, that continuation 
from  $N_f$ in the conformal window($4\leq N_f\leq 6$) to $N_f=0$ would be possible \cite{BroLu,BrGGR,Gru1}. 

We remark that the Orsay group analyzed  QCD running coupling in the Landau gauge from three gluon vertex. They separate the momentum space into $q<0.8 GeV$ and $1.5GeV<q$ and fit the lower momentum region by the instanton liquid model and 
 the higher momentum region by including $1/q^2$ power correction due to gluon
condensates \cite{orsay3}. We did not take $\alpha_s(0)=0$. Agreement of the
lattice results of QCD running coupling in $q>0.8$ GeV region and the three-loop perturbation theory is reported in  \cite{Lue}.

We observed that the contour-improved perturbation theory  performs a resummation of the perturbation series and that we can understand qualitatively the
Landau gauge lattice QCD data via these methods. 
\begin{acknowledgments}
We are grateful to Daniel Zwanziger for enlightenning discussions. S.F. thanks Kei-Ichi Kondo, Stanley Brodsky, Karel van Acoleyen, David Dudal and Kurt Langfeld for valuable information.

This work is supported by the KEK supercomputing project No. 03-94.
\end{acknowledgments}

\appendix*
\section{The numerical calculation of the Faddeev-Popov inverse}
In this appendix we briefly explain the numerical method of calculating the Faddeev-Popov inverse. 

\subsection{Perturbative method}
The ghost propagator, which is the Fourier transform of an expectation value of the inverse Faddeev-Popov operator ${\cal  M}=-\partial D=-\partial^2(I-M)$,
\begin{equation}
D_G^{ab}(x,y)=\langle {\rm tr} \langle \lambda^a x|({\cal  M}[U])^{-1}|
\lambda^b y\rangle \rangle,
\end{equation}
where the outmost $\langle\cdots \rangle$ denotes average over samples $U$, 
is evaluated as follows.  We take the plane wave for the source
${\bf b}^{[1]}={\bm\lambda}^b e^{iqx}$ and get the solution $-\Delta {\bm \phi}^{[1]}={\bf b}^{[1]}$. Here
${\bm \phi}^{[1]}=(-\Delta )^{-1}{\bf b}^{[1]}$. 
We calculate iteratively  $\phi^{[i+1]}=M{\bm \phi}^{[i]}({\bf x})$($i=1,\cdots,k-1$).
   The iteration was
continued until the maximum norm Max$_{\bf x}|{\bf \phi}^{[k]}({\bf x})|/$Max$_{\bf x}|\sum_{i=1}^{k-1}{\bm \phi}^{[i]}({\bf x})|<0.001\sim 0.01$.
The number of iterations $k$ is of the order of 60, in SU(2), $16^4$ lattice,  and of the order of 100 in SU(3).
We measure also the $\l_2$ norm $\||{\bf \phi}^{[k]}({\bf x})\||/\||\sum_{i=1}^{k-1}{\bm \phi}^{[i]}({\bf x})\||$.

We define ${\bm\Phi}^b({\bf x})=\sum_{i=1}^k{\bm \phi}^{[i]}({\bf x})$ and evaluate
$\langle \lambda^a e^{iqx},\Phi^b({\bf x})\rangle$ as the ghost propagator from
color b to color a.

In the low-momentum region of SU(2) we observed a specific color symmetry violation pattern, and in the case of SU(3) relatively large color off-diagonal matrix elements suppressed the color diagonal matrix element.  For a cross-check of the perturbative method, we adopted the straightforward conjugate gradient method and the 
preconditioned conjugate gradient method \cite{Ort} in which the truncated perturbation series is used for the preconditioning.

\subsection{Preconditioned conjugate gradient method}

We define ${\cal M}=-\partial^2(I-M)$ and define the truncated ${\cal M}^{-1}$
which is used in the perturbative method as $B^{-1}=(I+M+\cdots+M^{m-1})(-\Delta)^{-1}$.
First we choose ${\bf x}^0$ and define ${\bf r}^0={\bf b}-{\cal M}{\bf x}^0$.
Using the multigrid Poisson solver we calculate the perturbation series
\begin{equation}
{\tilde {\bf r}}^0=B^{-1}{\bf r}^0
\end{equation}
and define ${\bf p}^0={\tilde {\bf r}}^0$.

Then we begin the iteration for $k=0,1,\cdots$,
\begin{equation}
\alpha_k=-({\tilde {\bf r}}^k,{\bf r}^k)/({\bf p}^k,{\cal M}{\bf p}^k),
\end{equation}
\begin{equation}
{\bf x}^{k+1}={\bf x}^k-\alpha_k {\bf p}^k,
\end{equation}
\begin{equation}
{\bf r}^{k+1}={\bf r}^k+\alpha_k {\cal M}{\bf p}^k.
\end{equation}

We check the norm of ${\bf r}^{k+1}$, and if it is not small,
 we calculate the perturbation series ${\tilde {\bf r}^{k+1}}=B^{-1}{\bf r}^{k+1}$  as before. 
 We define
\begin{equation}
\beta_k=(\tilde{\bf r}^{k+1},{\bf r}^{k+1})/(\tilde{\bf r}^{k},{\bf r}^{k}),
\end{equation}
\begin{equation}
{\bf p}^{k+1}=\tilde{\bf r}^{k+1}+\beta_k{\bf p}^k,
\end{equation}
and go back to the beginning of the iteration cycle. By choosing a sufficiently
large number of $m$, the convergence occurs after a few iteration cycles.

The preconditioned method makes the $\l_2$ norm convergence faster than the
straightforward conjugate gradient method, but its maximum norm is larger than
that of straightforward method. The solution agrees with the straightforward conjugate gradient method within errors in the whole momentum region, but disagrees with the perturbative method in the lowest momentum point of $\beta=6.4$, SU(3) $48^4$ lattice.


\begin{thebibliography}{99}
\bibitem{KO} T. Kugo and I. Ojima, {Prog. Theor. Phys. Suppl.} {\bf 66}, 1 (1979).

\bibitem{Gr} V.N. Gribov, {\NPB} {\bf 139}{1}{(1978)}.

\bibitem{Zw} D. Zwanziger, {\NPB} {\bf 364} ,{127} {(1991)}, idem B
{\bf 412}, {657} (1994).

\bibitem{Zw1} D. Zwanziger, {\PRD}(to be published) ,hep-ph/0303028.

\bibitem{HdF} J.E. Hetrick and P.H. de Forcrand, {\NPB} (Proc. Suppl.){\bf 63A-C}, 838 {(1999)}.

\bibitem{Kond} K.I. Kondo, hep-th/0303251.

\bibitem{Bon} F.D.R. Bonnet et al., {\PRD}{\bf 64},{034501}{(2001)}.

\bibitem{Ni} M. Chaichian and K. Nishijima, hep-th/9909159 and references therein.


\bibitem{SHA} L. von Smekal, A. Hauck,  R. Alkofer, {Ann. Phys.} (N.Y.) {\bf 267},1 (1998).

\bibitem{Cu} A. Cucchieri and D. Zwanziger, {\PLB} {\bf 524},{123}{(2002)}.

\bibitem{orsay1} D. Becirevic et al., {\PRD} {\bf 61},{114508}{(2000)}.

\bibitem{adelade} D.B. Leinweber, J.I. Skullerud, A.G. Williams and C. Parrinello, {\PRD}{\bf 60},{094507}{(1999)}; ibid {\PRD}{\bf 61},{079901}{(2000)}.

\bibitem{NF} H.Nakajima and S. Furui, {\NPB} (Proc Suppl.){\bf 63A-C},{635, 865}(1999), 
 {\NPB} (Proc Suppl.){\bf 83-84},521 (2000), {\bf 119},730(2003);  
{\NPA} {\bf 680},{151c}(2000),
hep-lat/0006002, 0007001, 0208074.

\bibitem{FN} S. Furui and H. Nakajima, in {\it Quark Confinement and the Hadron Spetrum IV}, Ed. W. Lucha and K.M. Maung, World Scientific, Singapore, p.275(2002), hep-lat/0012017.
\bibitem{scgt} H. Nakajima and S. Furui, in {\it Strong Coupling Gauge Theories and Effective Field Theories}, Ed. M. Harada, Y. Kikukawa and K. Yamawaki, 
World Scientific, Singapore, p.67(2003), hep-lat/0303024.
\bibitem{lat03} H. Nakajima and S. Furui, Lattice '03 proceedings(2003), hep-lat/0309165.
\bibitem{CaMa} N. Cabibbo and E. Marinari, {\PLB}{\bf 119},{387}{(1982)}.
\bibitem{KePe} A.D.Kennedy and B.J.Pendleton, {\PLB}{\bf 156},{393}{(1985)}.
\bibitem{MN} T. Maskawa and H. Nakajima, {\PTP} {\bf 60},1526(1978),
{\PTP} {\bf 63},  641(1980).

\bibitem{SF} M. A. Semenov-Tyan-Shanskii and V. A. Franke, Zap. Nauchn.
Semin., LOMI {\bf 120} 159(1982).
\bibitem{Hack} W. Hackbusch, Multi-Grid Methods and Applications, Springer Series in Computational Mathematics 4, Springer, Berlin(1980).
\bibitem{Ort} J.M. Orthega, {\it Introduction to Parallel and Vector Solution of Linear Systems}, (Plenum) (New York) 1988), p.200.

\bibitem{PMS} P.M.Stevenson, {\PRD}{\bf 23}, {2916}{(1981)};

\bibitem{Gru} G. Grunberg, {\PRD}{\bf 29}, {2315}{(1984)};

\bibitem{HoMa} D.M. Howe and C.J. Maxwell, hep-ph/0204036 v2.

\bibitem{tHo} G. t'Hooft, in {\it The  Whys of Subnuclear Physics},ed. A. Zichichi, (Plenum, New York, 1978), pp.943-971.

\bibitem{vAc} K. van Acoleyen and H. Verschelde, {\PRD}{\bf 66},125012(2002),hep-ph/0203211.

\bibitem{ChRe} K.G. Chetyrkin and A. Retey, hep-ph/0007088.
\bibitem{BroLu} S.J. Brodsky and H.J. Lu, {\PRD}{\bf 51},{3652}{(1995)}
\bibitem{CeGo} W. Celmaster and R.J. Gonsalves, {\PRD}{\bf 20},{1420}{(1979)}.

\bibitem{orsay2} Ph. Boucaud et al., hep-ph/0003020 v2.

\bibitem{Bro} S.J. Brodsky, S. Menke and C. Merino, {\PRD}{\bf 67},{055008}{(2003)},hep-ph/0212078 v3.

\bibitem{FAR} C.S. Fischer, R. Alkofer and H. Reinhardt, hep-ph/0202195.

\bibitem{BCLM} J.R.C.Bloch, A. Cucchieri, K.Langfeld and T.Mendes, hep-lat/0209040 v2. 
\bibitem{AFF} C. Alexandrau, Ph. de Forcrand and E. Follena, hep-lat/0009003.
\bibitem{cuc} A. Cucchieri, {\NPB}{\bf508},{353}{(1997)}, hep-lat/9705005.
\bibitem{KoShi} K-I. Kondo and T. Shinohara, {\PLB}{\bf 491},{263}{(2000)}

\bibitem{Sha} M. Shaden, in {\it Quark Confinement and Hadron Spectrum VI},ed. W. Lucha and K.M Maung, (World Scientific, Singapore, 2002) p.258.
\bibitem{BM} I.L. Bogolubovsky and V.K Mitrjushkin, hep-lat/0204006 v2.
\bibitem{dudal} D.Dudal,H. Vershelde, V.E.R. Lemes, M.S. Sarandy, S.P.Sorella, M.Picariello, A. Vicini and J.A. Gracey, JHEP06(2003)003.
\bibitem{GGK} E. Gardi, G. Grunberg and M. Karliner, JHEP{\bf 07},{007}{(1998)},hep-ph/9806462
\bibitem{Mag} B.A. Magradze, hep-ph/0010070 v4.
\bibitem{Shir} D.V. Shirkov, hep-th/0210013 v3 and references therein.
\bibitem{BrGGR} S.J. Brodsky, E. Gardi, G. Grunberg and J.Rathsman, {\PRD} {\bf 63},{094017}{(2001)}. 
\bibitem{Gru1} G. Grunberg, {\PRD}{\bf 65},021701(R){(2001)}
\bibitem{orsay3} Ph. Boucaud et al., {\PRD}{\bf 63},{114003}{(2001)}; hep-ph/0212192.
\bibitem{Lue} M. L\"uscher, in {\it International Conference on Theoretical Physics, TH2002}, ed. by D. Iagolnitzer et al., (Birkh\"auser, Basel, 2003) pp.197-210, hep-ph/0211220.

\end{thebibliography}
\end{document}